\documentclass[floatfix,twocolumn,prb]{revtex4}
\usepackage{graphicx}
\usepackage{amsmath}
\usepackage{bm}
\usepackage{dcolumn}

\begin{document}

\title{Structurally-driven magnetic state transition of biatomic Fe
chains on Ir(001)}

\author{Yuriy~Mokrousov$^{1,2}$}
\email[corresp.\ author: ]{y.mokrousov@fz-juelich.de}
\author{Alexander~Thiess$^{1,2}$}
\author{Stefan~Heinze$^{1,3}$}
\affiliation{$^1$Institute of Applied Physics, University of Hamburg,
D-20355 Hamburg, Germany}
\affiliation{$^2$Institut f\"ur Festk\"orperforschung and
Institute for Advanced Simulation,
Forschungszentrum J\"ulich, D-52425 J\"ulich, Germany}
\affiliation{$^3$Institute of Theoretical Physics and Astrophysics,
Christian-Albrechts-University of Kiel, D-24098 Kiel, Germany}
\date{\today}

\begin{abstract}

Using first-principles calculations, we demonstrate that the
magnetic exchange interaction and the magnetocrystalline
anisotropy of biatomic Fe chains grown in the trenches of the
$(5\times1)$ reconstructed Ir(001) surface depend sensitively
on the atomic arrangement of the Fe atoms. Two structural
configurations have been considered which are suggested from
recent experiments. They differ by the local symmetry and the
spacing between the two strands of the biatomic Fe chain.
Since both configurations are very close in total energy they
may coexist in experiment. We have investigated collinear ferro-
and antiferromagnetic solutions as well as a collinear state
with two moments in one direction and one in the opposite
direction ($\uparrow \downarrow \uparrow$-state). For the
structure with a small interchain spacing, there is a strong
exchange interaction between the strands and the ferromagnetic
state is energetically favorable. In the structure with larger
spacing, the two strands are magnetically nearly decoupled
and exhibit antiferromagnetic order along the chain. In both
cases, due to hybridization with the Ir substrate the exchange
interaction along the chain axis is relatively small compared to
freestanding biatomic iron chains. The easy magnetization axis
of the Fe chains also switches with the structural configuration
and is out-of-plane for the ferromagnetic chains with small
spacing and along the chain axis for the antiferromagnetic
chains with large spacing between the two strands. Calculated
scanning tunneling microscopy images and spectra suggest the
possibility to experimentally distinguish between the two
structural and magnetic configurations.
\end{abstract}

\maketitle

\section{Introduction}

Driven by the wish to realize the proposed concepts of
future spintronic devices~\cite{Prinz1998,Wolf2001,Fert2007}
the development of novel nanostructures and nanomaterials
with tailored electronic and magnetic properties has become
a key challenge of today's research. A promising path to control
the magnetic properties of matter is to use low-dimensional systems
and to reduce their size down to the nanometer or even atomic
scale. For nanoscale systems, however, an essential requirement
is to enhance the magnetic anisotropy in order to stabilize
magnetic order against thermal fluctuations or quantum tunneling.
The manipulation of exchange interactions opens another path to
create new materials with a magnetic state that may be tunable
by external magnetic or electric fields.\cite{Tsymbal} E.g.~the occurrence of
chiral spin spiral states at surfaces has been
demonstrated,~\cite{Bode2007,Ferriani2008} their manipulation
by electrical currents was suggested~\cite{Bode2007,Pfleiderer2007}
and a way to grow films with multiple metastable magnetic
states has been proposed.~\cite{Ferriani2007}

The ability to create one-dimensional monoatomic magnetic chains of
transition-metals on surfaces by self-organization
\cite{Gambardella2002,Gambardella2004} or by manipulation with a
scanning tunneling microscope \cite{Hirijebedin2006} has recently
opened new vistas to explore and manipulate artificial magnetic
nanostructures even atom-by-atom. E.g.~the pioneering work of
Gambardella {\it et al.}\cite{Gambardella2002,Gambardella2004}
demonstrated that the magnetic anisotropy of atomic
transition-metal chains, consisting of Co atoms on a stepped
Pt(111) surface, is dramatically enhanced with decreasing
dimensionality from two-dimensional films to quasi-one-dimensional
chains and depends sensitively on the number of Co strands
in a chain. These experimental observations were explained based
on electronic structure calculations which emphasized the crucial
role played by the substrate, reduced symmetry, and structural
relaxations for the magneto-crystalline
anisotropy.\cite{Dorantes1998,Komelj2002,Ujfalussy2004,Baud2006}
The large magnetic anisotropies led to slow
relaxation dynamics of the magnetization and the observation of
magnetic hysteresis loops at low temperatures indicative of
ferromagnetic coupling. In another experiment, Mn chains of up
to 10 atoms were created by manipulation with a scanning tunneling
microscopy tip on an insulating CuN layer grown on Cu(001). Experimentally,
the exchange interaction between individual spins was obtained 
by measuring the excitation spectrum via inelastic tunneling
spectroscopy which showed the quantum behavior of the
entire chain.\cite{Hirijebedin2006} Even the sign and size of
the exchange interaction between the Mn atoms could be extracted
from the experimental data. Calculations based on
density-functional theory (DFT) clarified that a superexchange
mechanism along the Mn-N-Mn bond is responsible for the weak
antiferromagnetic coupling.\cite{Rudenko2009,Urdaniz}

Recently, Hammer {\it et al.}~have used a combination of IV-LEED
and scanning tunneling microscopy (STM) measurements to demonstrate
that the Ir(001) surface can serve as an ideal template to
grow defect-free, nanometer long transition-metal nanowires of
different structure, chemical composition, and length depending
on the preparation
conditions.\cite{Hammer2003a,Hammer2003b,Klein2004}~E.g.~biatomic
Fe chains can be created on the $(5\times1)$ reconstructed Ir(001)
surface and lifting the surface reconstruction by hydrogen opens
the possibility to produce Fe-Ir-Fe triatomic chains. While the
biatomic Fe chains are a very promising system to study magnetism
of (quasi-) one-dimensional transition-metal chains, there is
little understanding so far. Experimentally, it is extremely
challenging as for laterally averaging measurements samples with
a homogeneous distribution of chains are needed or a technique must
be applied which allows to locally probe the magnetic properties
of individual Fe chains. Another key difficulty is the detailed
characterization of the chains structure. From combined STM and LEED
experiments it is only known that the biatomic chains grow in the
trenches of the $(5\times1)$ reconstructed Ir(001) surface, however,
the adsorption sites in the trenches could not be deduced. As
structure and magnetism are closely correlated in such systems,
their magnetic properties are an open issue.

A theoretical study using first-principles calculations~\cite{Spisak2003}
reported an excellent agreement with the structural parameters of
the $(5\times1)$ reconstructed Ir(001) surface and concluded that
the Fe chains are strongly ferromagnetic at low temperatures but
were probably non-magnetic in the room temperature measurements
of Hammer {\it et al.}~However, this theoretical study considered
only ferromagnetic solutions and did not determine the magnetocrystalline
anisotropy energy which is crucial for an experimental verification
of the proposed ferromagnetism in these chains. On the other hand, a
$5d$ transition-metal substrate such as Ir can have dramatic consequences
on the exchange coupling in a deposited Fe nanostructure as is apparent
from the observation of a complex nanoscale magnetic structure for an Fe
monolayer on Ir(111),\cite{vonBergmann2006} the antiferromagnetic ground
state of an Fe monolayer on W(001),\cite{Kubetzka2005} to name just a
few. Recent studies on Fe stripes on Pt(997)\cite{Honolka2009:1} and
FePt surface alloys\cite{Honolka2009:2} report on strong correlation
between complex magnetic ground states and the details of structural
arrangement.

\begin{figure*}
\begin{center}
\hspace{1.0cm}  \includegraphics[width=0.85\textwidth]{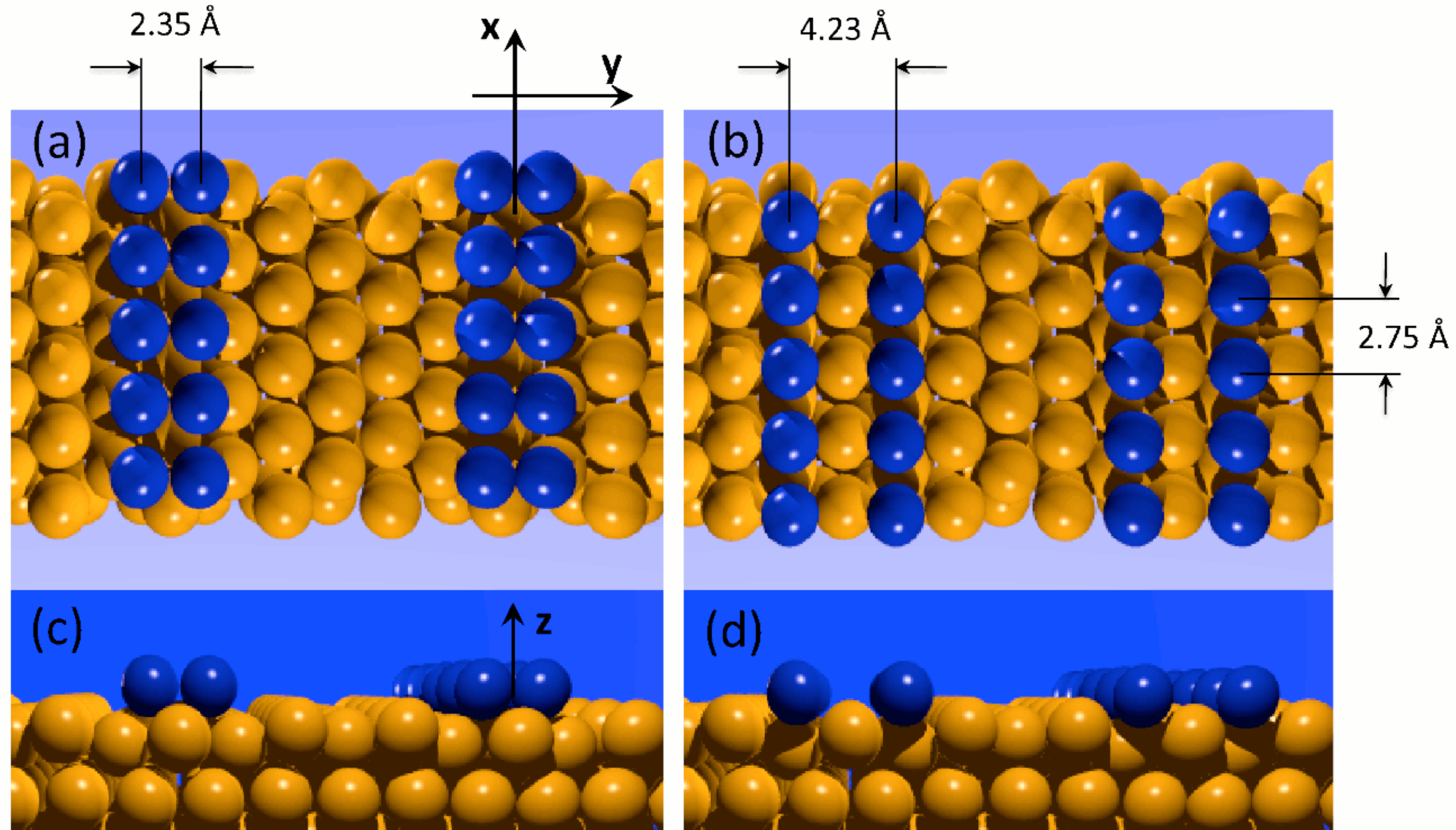}
\includegraphics[width=0.365\textwidth]{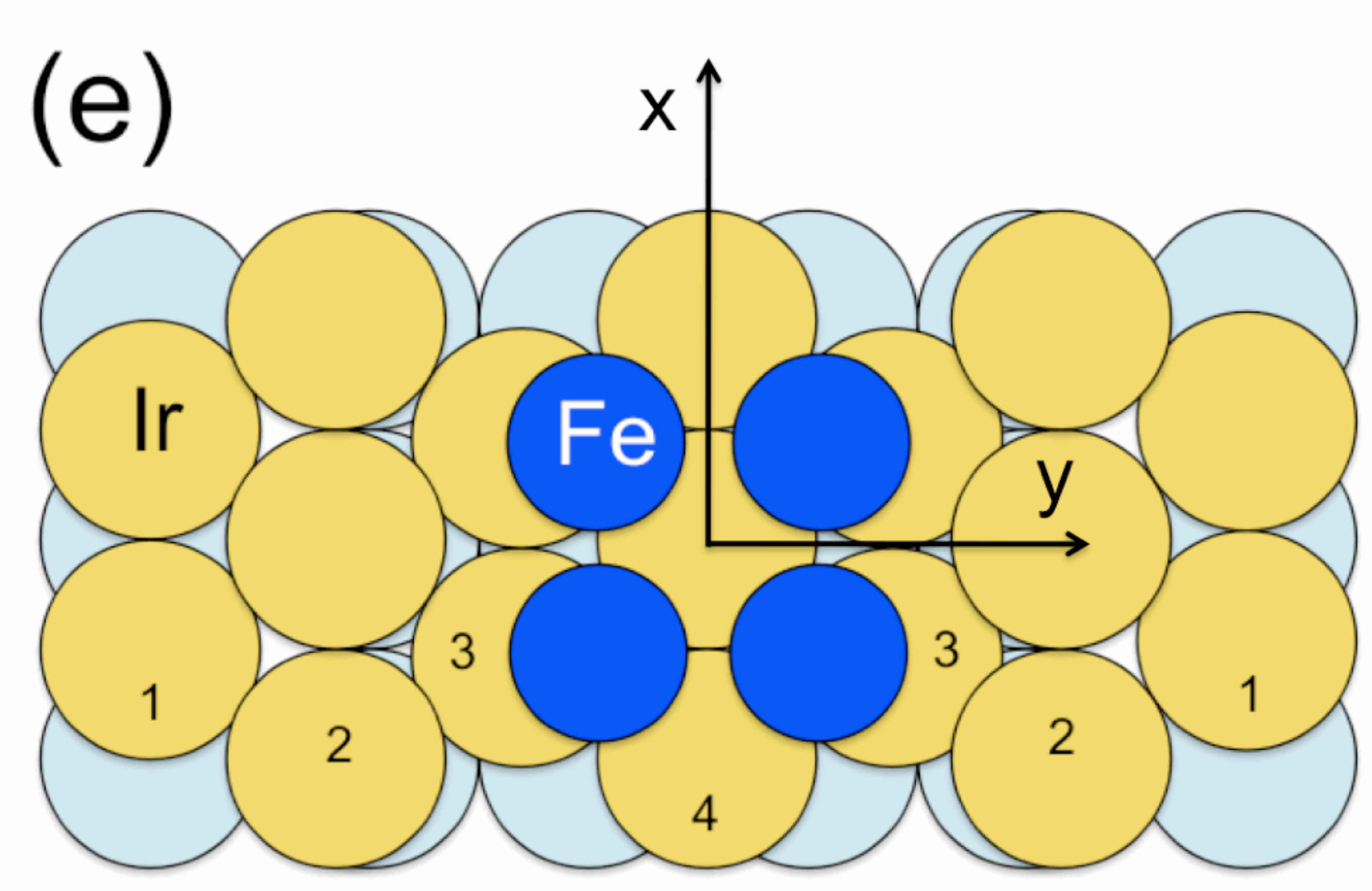}$\quad\quad$
\includegraphics[width=0.365\textwidth]{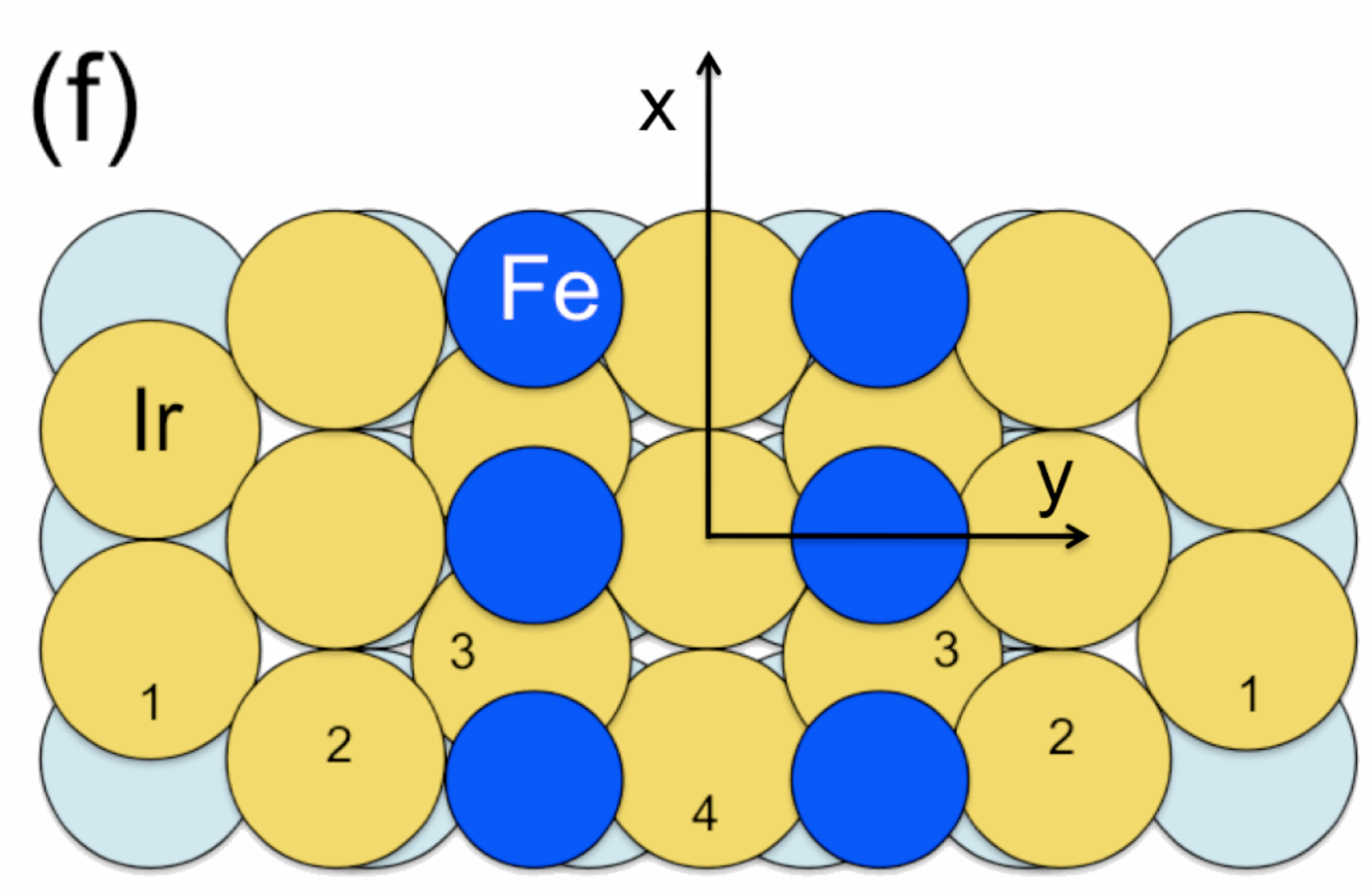}
\end{center}
\caption{\label{fig1} (color online) Geometrical structure of the
biatomic Fe chains on the $(5\times1)$ reconstructed Ir(001)
substrate in C1 ((a)-top view, (c)-side view) and C4 ((b)-top
view, (d)-side view) configurations. Fe atoms are marked in blue,
while Ir atoms are marked in gold. In (e) and (f) a schematic top
view of respectively C1 and C4 biatomic Fe chains is shown
(positions of the atoms do not correspond to realistic
calculated values). In the latter graphs the numbering of the Ir atoms
corresponds to that in Fig.~2 and table~II. The $x$- and $y$-axis defined 
in (a) correspond to [011] and [0\={1}1] directions, respectively.}
\end{figure*}

Here, we use first-principles calculations based on density-functional
theory to study the structural, electronic, and magnetic properties of
biatomic Fe chains deposited in the trenches of the $(5\times1)$
reconstructed Ir(001) surface.\cite{footnote} We focus on two 
structural arrangements of the biatomic chains which differ by the
adsorption sites of the Fe atoms.  In one configuration the distance
between the two strands of the biatomic chain is smaller than the atom
spacing along the chain direction (denoted as C1 in accordance with
Ref.~\onlinecite{Spisak2003}), while in the other their separation is
clearly larger than the interchain spacing (denoted as C4),
see Fig.~\ref{fig1}. We consider collinear ferro- and antiferromagnetic
arrangements along the chains and the $\uparrow \downarrow \uparrow$-state
with two moments in one and one moment in the opposite direction.
We find that the energetically favorable magnetic state as well as
the easy magnetization axis of the Fe chains depend sensitively on
their atomic arrangement and local symmetry. For the Fe chains in the
C1-structure, the exchange coupling along the chain is ferromagnetic,
however, due to hybridization with the Ir substrate it is much weaker
than for freestanding biatomic Fe chains with the same atom spacing.
Due to their small
separation, the two strands of the C1-chains are also ferromagnetically coupled.
Surprisingly, for the C4-configuration, we find a transition from ferromagnetic
state for free-standing chains to antiferromagnetic order along the biatomic chains
after deposition on the Ir substrate. In this case, the hybridization with the
Ir surface is strong enough to invert the sign of exchange coupling, while the
two Fe strands are nearly exchange decoupled. The interplay of the Fe
interstrand distance and the hybridization with the Ir substrate results also
in a different easy axis of the magnetization for the two structures: while in
the C1-FM state the easy axis is out-of-plane, it switches into the chain axis
for the C4-AFM configuration. The total energy of the C1- and
C4-structure are quite close and therefore both chain types could occur
in an experiment depending on the growth conditions. We simulate measurements
by STM and observe that the two strands of the biatomic Fe chain in the
C1-ferromagnetic state are too close to be individually resolved, while
they can be distinguished in the C4-structure with an antiferromagnetic
ground state. In the latter case, spin-polarized STM (SP-STM) should
further allow to directly resolve the two-fold magnetic periodicity
along the chain.

The structure of this paper is as follows. In the next section we
present details of our method and the calculations. Then we discuss
the structural relaxations of the pure $(5\times1)$ reconstructed Ir(001)
substrate and of the biatomic Fe chains in the two structural configurations
on the $(5\times1)$ reconstructed Ir(001) surface. In section~IV, we
analyze the magnetic ground state configuration and the effects of
hybridization with the Ir substrate, before we turn to the magnetocrystalline
anisotropy in section~V. We explore the feasibility to experimentally
resolve the different structural and magnetic properties by SP-STM and
to verify our predictions of a structure-dependent magnetic ground state.
Finally, a conclusion and summary is given.

\section{Computational details}

We employed the film-version of the
full-potential linearized augmented plane-wave (FLAPW) method, as
implemented in the J\"ulich density-functional theory (DFT) code
{\tt FLEUR}. We used inversion-symmetric films with 7 and 9
layers of the $(5\times 1)$ reconstructed Ir(001) surface and Fe
biatomic chains on both sides of the slab. The whole system
possesses spatial inversion symmetry. We calculate the biatomic
Fe chains in a $(5\times 1)$ supercell along the $y$-direction
([0\={1}1]-axis), perpendicular to the chain $x$-axis ([011]-axis), 
which results in a distance of 13.75~\AA~between the axes of two
adjacent biatomic chains, c.f.~Fig.~\ref{fig1} for structural
arrangements and definition of the axes. The theoretical Ir lattice
constant we used for calculations constituted a value of 3.89~\AA.
We used the generalized-gradient approximation (GGA, revPBE 
functional\cite{revpbe}) of the
exchange-correlation potential for the structural relaxations and
tested total energy differences between the two magnetic
configurations also within the local-density approximation (LDA, VWN 
functional\cite{vwn}).
We used 18 $k$-points in a quarter of the full two-dimensional
Brillouin zone (2D-BZ) for self-consistent calculations. The
calculated total energy differences between different magnetic
ground states were carefully tested with respect to the number
of $k$-points. For the basis functions, we used a cut-off parameter
of $k_{\rm max}=3.6$~a.u.$^{-1}$ for relaxations and 3.7~a.u.$^{-1}$
for comparing the total energies of different magnetic
configurations.

We considered two possible structural arrangements of the Fe
atoms denoted as C1 and C4 according to the notation of
Ref.~\onlinecite{Spisak2003}, which are shown in Fig.~\ref{fig1}.
Experimentally, it has been observed by STM
\cite{Hammer2003a} that the biatomic chains grow in
the trenches of the (5$\times$1) reconstructed Ir(001) surface
and we therefore focus on these two configurations.
Relaxations were performed until the forces changed by less than
$3\cdot 10^{-4}$~htr/a.u. The convergence of the relaxed atomic
positions was carefully tested with respect to the computational
parameters. Relaxations were performed only for the ferromagnetic
state of both C1 and C4 configurations and the antiferromagnetic
and the $\uparrow \downarrow \uparrow$-states were calculated
on these atomic positions.

\section{Structure and Relaxations}

As a first step, we performed a structural relaxation of the pure
$(5\times1)$ reconstructed Ir(001) surface. As can be seen in
table \ref{Table_relaxation}, the results we obtain with a film
of 9 layers agree very well with the experimental data measured
by IV-LEED.~\cite{Lerch2006} In particular, the experimentally
observed trench-like structure is reproduced. Our values are also
in close agreement with those obtained with the VASP code by
Spi\v{s}\'ak {\it et al.}\cite{Spisak2003} For deposited biatomic
Fe chains we have also found a very good agreement between the
values of the relative atomic positions obtained by relaxing a
slab with 9 and 7 layers of the Ir substrate
(see table~\ref{Table_relaxation}).

\begin{figure}
\begin{center}
   \includegraphics[scale=0.12]{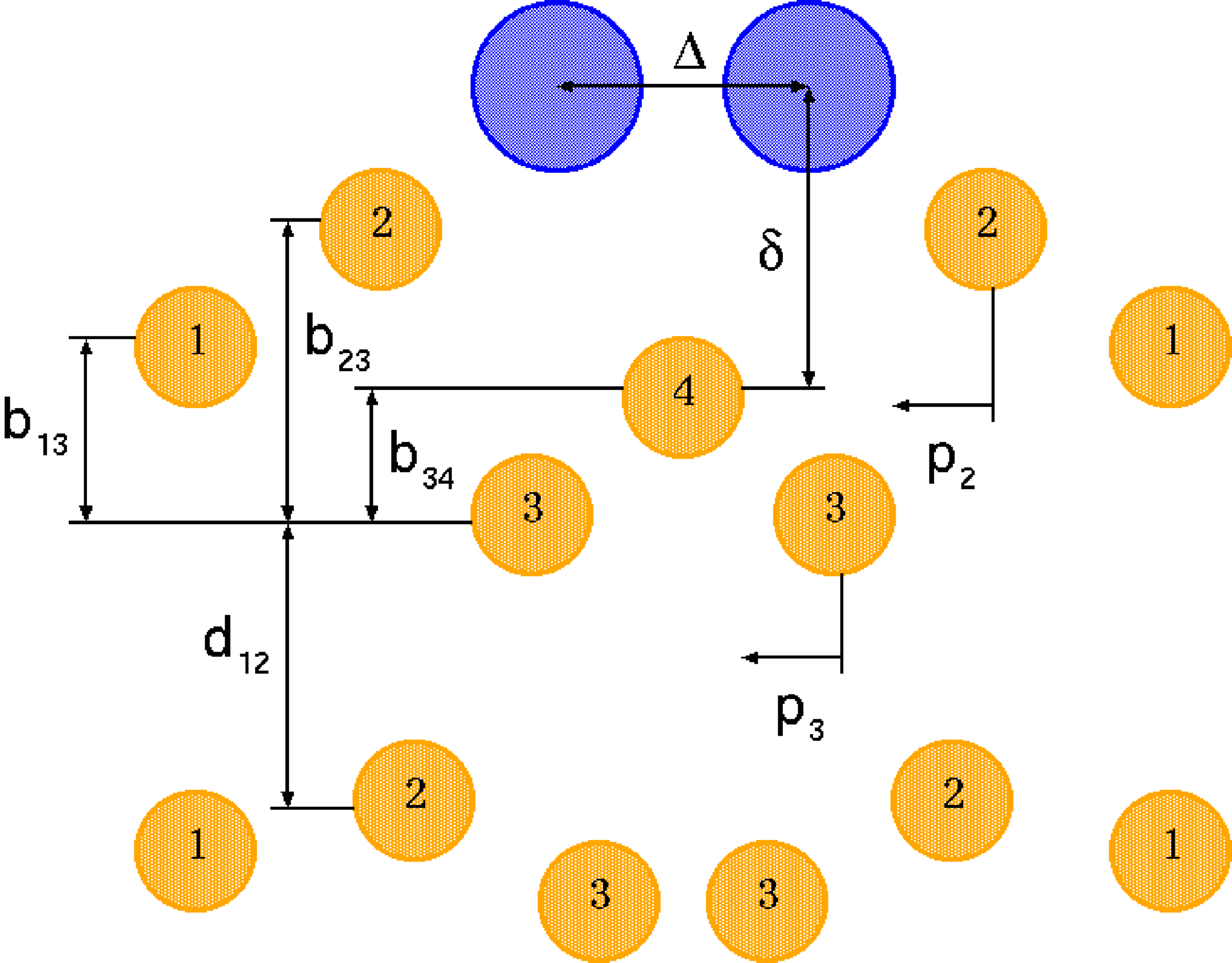}
\end{center}
\caption{\label{fig2} (color online)
 Schematical representation of the structural parameters given in
 table I. Note that in this figure the buckling of the Ir substrate
 is greatly exaggerated and the positions of the Fe atoms (blue) and
 Ir atoms in the first and second layer of the substrate (orange)
 do not correspond to realistic calculated values.
 \label{fig:sketch}
}
\end{figure}

We performed relaxations of the Fe biatomic chains only for the
ferromagnetic solutions to reduce the computational effort,
presenting the results in table~\ref{Table_relaxation}. While
the distance $\Delta$ between the atomic strands in
C1-configuration is 2.35~\AA, and thus smaller than along the
chain (see Figs.~1~and~2), it is almost twice larger
in the C4-configuration, i.e.~$\Delta=4.23$~{\AA}, and the two
strands are well separated. In the latter structure, the Fe atoms
are also more embedded into the Ir surface which is illustrated by
the smaller vertical distance $\delta$ from the Ir surface atoms
between the two Fe chain atoms (denoted as Ir4 in
Fig.~\ref{fig:sketch}). The influence of the C4-chains on the
buckling of the Ir substrate is also more pronounced as can be
seen from the increased vertical separations of the surface Ir
atoms. The different structural relaxations of the two chain
configurations already hint at a larger hybridization of the Fe
$3d$ and Ir $5d$ states and a stronger influence in the C4
arrangement. For both configurations the distance between the
Fe dimers along the chain's axis was imposed by the Ir substrate
and constituted 2.75~\AA.

\begin{table}
\begin{ruledtabular}
\begin{tabular}{cccccc}
          &  LEED    &  $(5\times1)$ Ir & $(5\times1)$ Ir   &  C1-FM  &  C4-FM \\ \hline
9 layers  &   Ref.~\onlinecite{Lerch2006}       &    Ref.~\onlinecite{Spisak2003}   &          &        &        \\ \hline
$\Delta$  &   $-$    &          &          &  2.35   &  4.23  \\
          &          &          &          & (2.65)  & (4.35) \\
$\delta$  &   $-$    &          &          &  1.90   &  1.53  \\
$d_{12}$  &  1.94    &   2.00   &   1.97   &   1.93   &  1.82  \\
$b_{13}$  &  0.25    &   0.20   &   0.20   &  0.22   &  0.42  \\
$b_{23}$  &  0.55    &   0.47   &   0.54   &   0.62   &  0.77  \\
$b_{34}$  &  0.20    &   0.17   &   0.18   &  0.30   &  0.55  \\
$p_2$     &  0.05    &   0.03   &  0.04   &  0.07   &  0.04  \\
$p_3$     &  0.07    &   0.07   &   0.07   &  0.12   &  0.11  \\ \hline
7 layers  &          &          &          &         &        \\ \hline
$\Delta$  &   $-$    &          &          &  2.40   &  4.19  \\
$\delta$  &   $-$    &          &          &  1.90   &  1.54  \\
$d_{12}$  &  1.94    &   $-$   &   $-$    &  1.93   &  1.81  \\
$b_{13}$  &  0.25    &   $-$   &   $-$    &  0.22   &  0.42  \\
$b_{23}$  &  0.55    &   $-$   &   $-$    &  0.63   &  0.79  \\
$b_{34}$  &  0.20    &   $-$   &   $-$    &  0.30   &  0.53  \\
$p_2$     &  0.05    &   $-$   &   $-$    &  0.07   &  0.05  \\
$p_3$     &  0.07    &   $-$   &   $-$    &  0.13   &  0.12  \\
\end{tabular}
\end{ruledtabular}
\caption{\label{Table_relaxation}
 Relaxations of the Fe biatomic chains and the uppermost layer of the
 $(5\times1)$ reconstructed Ir(001) substrate for 9- and 7-layer slabs.
 All values are given in {\AA}ngstrom.
 The distances $d,b,$ and $p$ correspond to those in
 Ref.~\onlinecite{Lerch2006} and are depicted in Fig.~\ref{fig:sketch}.
 For comparison the relaxations of the bare Ir(001) substrate
 from Ref.~\onlinecite{Lerch2006} (IV-LEED data) and
 Ref.~\onlinecite{Spisak2003} are given. For the interchain distance
 $\Delta$ values is brackets correspond to those calculated
 in Ref.~\onlinecite{Spisak2003}.
}
\end{table}

\section{Magnetic order and exchange interactions}

Now we study the magnetic order and exchange interaction of the chains
in the two structural arrangements. We considered the ferro- (FM)
and antiferromagnetic (AFM) solution (with antiparallel magnetic moments
between adjacent Fe atoms along the chain) for both types of chains.
For comparison, we have also calculated free-standing biatomic Fe
chains with the same interatomic spacings as in the C4 and C1 
configurations. As can be seen in
table~\ref{Table_magnetism}, the magnetic moments of the Fe atoms are
slightly larger in the C4 configuration where the Fe atoms are further
apart. The difference in the magnetic moments between FM and AFM state
is very small. The induced moments in the Ir surface layer depend
much more sensitively on the magnetic state of the Fe chains. For the
FM solutions, these moments are significant in both structures and decay
slightly faster for the C1- than for the C4-configuration. In the AFM
state, due to symmetry some of the Ir atoms do not carry an induced
moment. In the C4 structure, the Ir surface atom between the Fe atoms
has a rather large moment of 0.18~$\mu_B$ indicating a strong
hybridization.

The total energy differences between the two magnetic configurations
given in table~\ref{Table_magnetism} reveal a surprising result. In
particular, we find that the preferred magnetic state is FM for the
Fe chains in the C1 structure ($d_{\rm{Fe-Fe}}$=2.35~\AA), while the
total energy difference is in favor of the AFM solution for the C4
structure with Fe atoms further apart ($d_{\rm{Fe-Fe}}$=4.23~\AA).
For free-standing biatomic Fe wires with the same spacing
between Fe atoms, the energy
difference between FM and AFM states is 74~meV/Fe and 164~meV/Fe
in favor of the FM state for the C1 and C4 configuration, respectively.
For supported chains the corresponding energy differences are of the 
order of rather small 30 meV/Fe, indicating a strong influence of the
Ir substrate on the magnetic coupling and weakening of the FM 
interaction in the free-standing chains due to hybridization with 
the Ir atoms.\cite{Mokrousov:supple} Because of the large separation
of the two strands of the biatomic chain in the C4 configuration,
this effect is dramatic and leads to the AFM ground state. This notion
is further supported by the magnitude of the exchange interaction 
between the two strands. We have calculated the total energy difference
between a FM and AFM alignment of the two strands of the deposited
Fe chains which is 166~meV/Fe and 4~meV/Fe in favor of ferromagnetic 
coupling in the C1- and C4-configuration, respectively. The extremely
small value in the C4-configuration, indicating that the two strands
are magnetically nearly decoupled, is probably due to a small
indirect exchange interaction via the Ir substrate.

We have also calculated the FM and AFM total energy differences for
both C1 and C4 configurations using 7 layers of Ir substrate, and
the results are compared in table~\ref{Table_magnetism} to those
obtained with 9 layers. In the C1-structure the FM-AFM energy
difference of 20.6~meV/Fe is in good agreement with the value
of 21.1~meV/Fe for 9-layers of substrate. In the C4-configuration,
on the other hand, the energy difference between the AFM ground
state and the FM state increases by only 7~meV/Fe for the
thinner substrate. These results show that the favorable magnetic
state within each structural arrangement is independent of the
substrate thickness.
However, in case of
9 layers the C4-AFM state is lower in energy than the C1-FM solution
by 9.8~meV/Fe, while in case of 7 layers this energy difference reverses
sign and constitutes 8.5~meV/Fe in favor of C1-FM state (see
table~\ref{Table_magnetism}). This discrepancy probably arises due to quantum
well states in the Ir substrate which do not influence the total
energy difference between different magnetic states within the
same structural arrangement. Overall, our calculations reveal
that, judging only from this energy difference, both configurations
might appear in experiment and can be observed via,~e.g., scanning
tunneling microscopy measurements.

We have further checked the influence of the exchange-correlation
potential on the FM-AFM energy differences and found very similar
results within the local-density approximation LDA. Using the
LDA, a 7 layer $(5\times1)$ Ir(001) substrate and the relaxed atomic
positions found with GGA, in the C1 configuration the FM state
is by 25.1~meV/Fe lower in energy than the AFM state (c.f.~a
value of 20.6~meV/Fe in GGA). In the C4 structure the AFM state
is by 23.7~meV/Fe lower than the FM state (c.f.~a value of
40.3~meV/Fe in GGA).

Within GGA, we also performed calculations of a collinear magnetic state with
a spin arrangement of $\uparrow\downarrow\uparrow$ (periodically
repeated along the chain axis) for the Fe biatomic chains on 7
layers of Ir substrate in the C1 and C4 structure. Our calculations
reveal that the C4-$\uparrow\downarrow\uparrow$ state is by
16.0~meV/Fe higher in energy than the C4-AFM state and by
24.3~meV/Fe lower in energy than the C4-FM state. In the C1
configuration, the $\uparrow\downarrow\uparrow$-state is 9.4~meV/Fe
higher in energy than the FM state, and by 11.2~meV/Fe lower in
energy than the AFM state. From these three collinear magnetic
solutions we can estimate the nearest-neighbor and next-nearest
neighbor exchange constants of an effective 1D Heisenberg model
to be $J_1=-10$~meV and $J_2=+1$~meV for the C4 structure and
$J_1=+5$~meV and $J_2=-1.3$~meV for the C1 structure. These
values illustrate the strong tendency towards antiferromagnetic
coupling due to the Ir substrate. Even in the C1-configuration,
the FM nearest-neighbor coupling has become very weak. Note, that
a similar influence of the Ir surface has recently been reported
for Fe monolayers on Ir(111) and for other
$4d$/$5d$ substrates.\cite{Hardrat2009}

\begin{figure*}
\begin{center}
\includegraphics[scale=0.7]{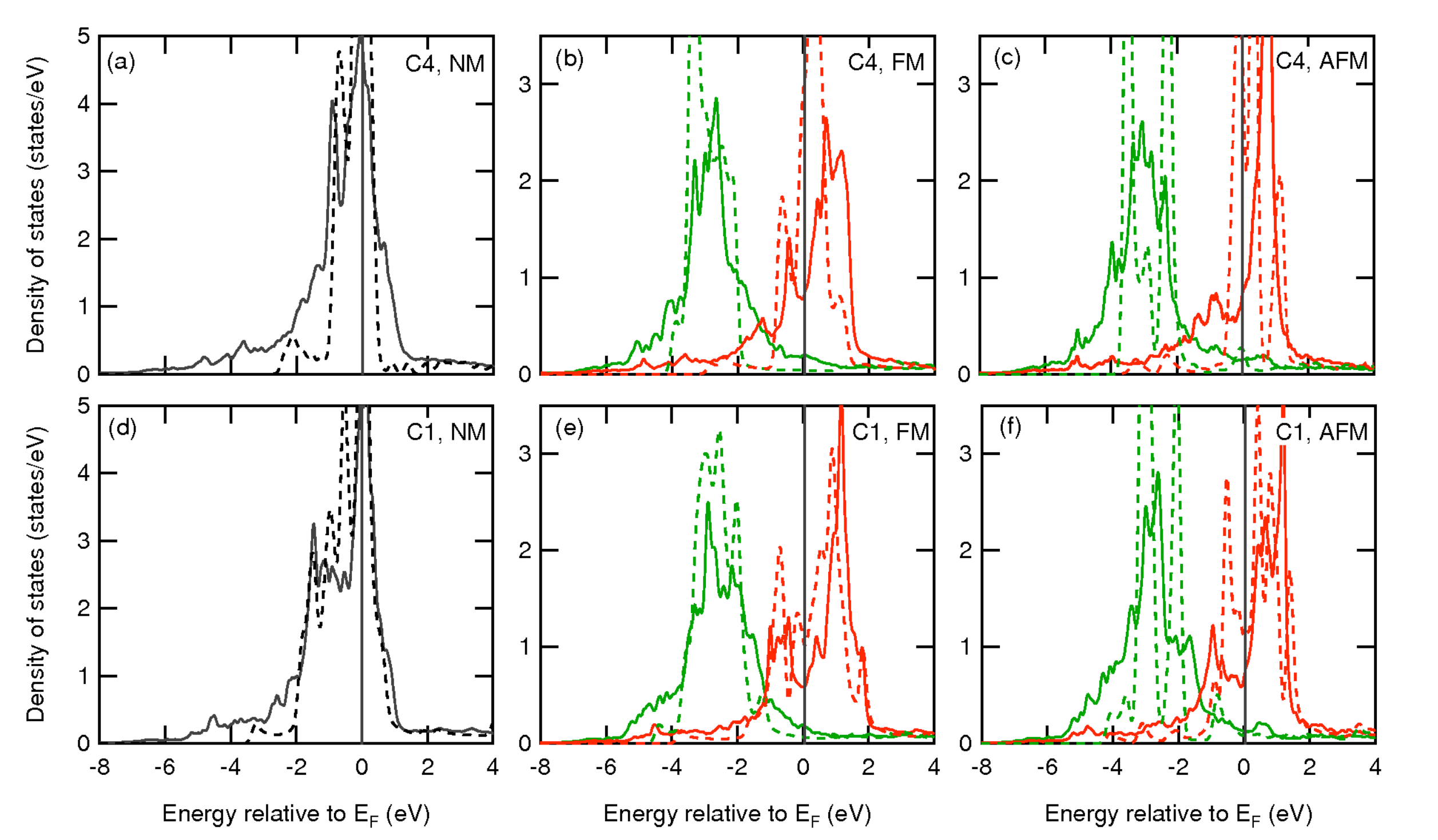}
\end{center}
\caption{\label{fig:MT-DOS} (color online) Density of states (DOS) for the 
nonmagnetic (NM), ferromagnetic (FM) and antiferromagnetic (AFM) state of Fe 
biatomic chains on 9 layers of the $(5\times1)$ reconstructed Ir(001) surface 
in C4, (a)-(c), and C1, (d)-(f), configuration. The DOS is given inside the 
muffin-tin spheres of Fe atoms. The dashed lines show the DOS for unsupported Fe 
biatomic chains with corresponding spacings between the Fe atoms in the 
chain. In (b), (c), (e) and (f) left (green) and right (red) curves stand
for spin-up and spin-down channels, respectively.}
\end{figure*}

In order to understand the sensitive dependence of the magnetic
coupling in the Fe chains upon the structural arrangement we
take a look at the density of states (DOS) for the two
configurations, shown in Fig.~\ref{fig:MT-DOS} in comparison
with the free-standing chains. In the non-magnetic state, the
DOS of the supported Fe chains displays a large peak at the
Fermi energy for both structures, however, the $d$-band width
is smaller in the C4 structure due to the larger separation
and weaker hybridization between the Fe atoms $-$ an effect
even more pronounced in free-standing wires.

\begin{figure}
\begin{center}
\includegraphics[scale=0.6]{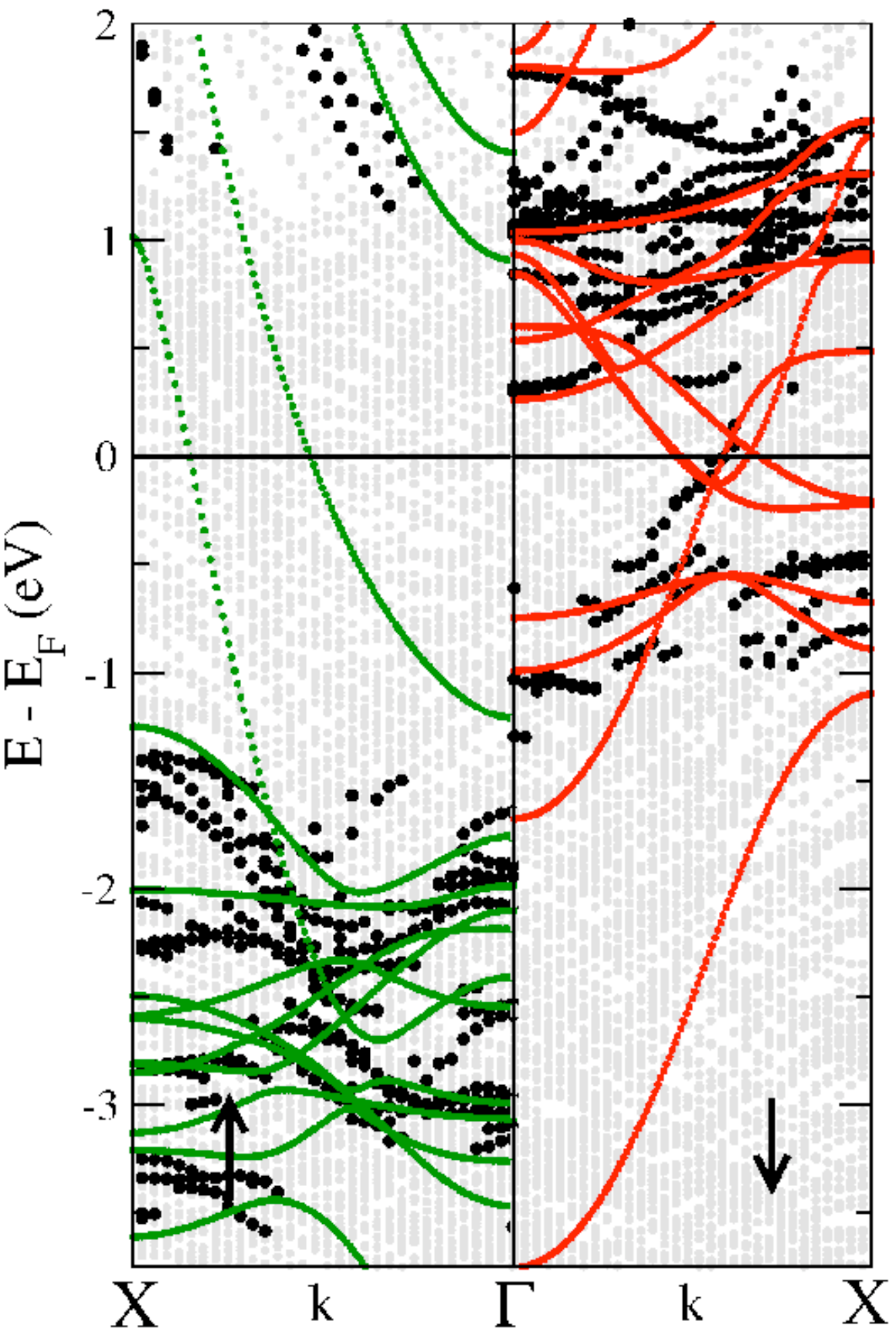}
\end{center}
\caption{\label{fig:BS_C1_FM} (color online) Bandstructure of the Fe
biatomic chains in the C1-FM configuration on 9 layers of the $(5\times1)$
reconstructed Ir(001) surface. Left panel shows majority bands and right
panel the minority bands. $k$-vector has been chosen along the chain direction. 
Large black filled circles denote states which are localized on the
Fe chain. For comparison the bands for an unsupported biatomic Fe chain
with the same interatomic spacing are given by green (majority) and red
(minority) filled small circles. Grey circles on the background mark the states
of the whole Fe+Ir system.}
\end{figure}

In the C1 structure, the direct interaction between the Fe atoms
perpendicular to the chain is much stronger than in C4-wires and
the free-standing chains are a better approximation. Correspondingly,
the FM DOS of the free-standing chains in C1-geometry is very similar
to the supported chains, while larger changes are visible in the C4-FM
configuration. This becomes even more evident from the comparison
of the bandstructures of the C1-FM free-standing and C1-FM supported
biatomic Fe chains presented in Fig.~\ref{fig:BS_C1_FM}. In this plot
the electronic states of the free-standing C1 chain (small red and 
small green circles) display a close correspondence to the states
of the C1 supported chain (large black circles) which are localized 
mainly inside the muffin-tin spheres of Fe atoms. Remarkably, for
many of the bands of the two systems a direct correspondence
in terms of symmetry can be made.

For the AFM solution in both structural configurations the modifications in the
DOS due to interaction with the substrate are quite significant.
The electronic bands in free-standing AFM chains are normally very
flat and corresponding peaks in the DOS are very
sharp\cite{Mokrousov:supple} $-$ in this case the effect of the
hybridization of the localized $3d$-orbitals with extended states
of Ir atoms on the DOS can be very strong. For both FM and AFM
magnetic states, a slightly larger exchange splitting can be observed
in C4 deposited chains, as compared to the C1 configuration, which
leads also to larger spin moments of Fe atoms in C4 arrangement
(c.f.~table~\ref{Table_magnetism}). Overall, an interplay of
decreasing Fe-Fe hybridization with increasing Fe-Ir hybridization
when going from C1 to C4 structural configuration leads to somewhat
larger localization of Fe electronic states in C4 biatomic chains.

\begin{table}
\begin{ruledtabular}
\begin{tabular}{ccccc}
          &  C1-FM   &  C1-AFM  &  C4-FM  & C4-AFM \\ \hline
UBC       &   3.12   &  3.14    &   3.28  &  3.22   \\ \hline
Energy   &  0       &  +74.1    &   +751.1     &  +914.6 \\ \hline\hline
9 layers  &          &          &         &        \\ \hline
Energy    &   +9.8     &   +30.9  & +33.7   &  0  \\ \hline
Total     &   3.21   &   0.00   &  3.41   &  0.00  \\
Fe        &   2.97   &   2.98   &  3.01   &  3.04  \\
Ir1       &   0.00   &   0.01   &  0.00   &  0.00  \\
Ir2       &$-$0.02   &   0.00   &  0.07   &  0.06  \\
Ir3       &   0.08   &   0.06   &  0.14   &  0.00  \\
Ir4       &   0.29   &   0.00   &  0.16   &  0.18  \\ \hline
7 layers  &          &          &         &        \\ \hline
Energy    &   0      &  +20.6   & +48.8   &  +8.5 \\ \hline
Total     &   3.23   &   0.00   &  3.34   &  0.00  \\
Fe        &   2.98   &   2.98   &  3.01   &  3.04  \\
Ir1       &   0.00   &   0.00   &  0.00   &  0.00  \\
Ir2       &$-$0.03   &   0.00   &  0.08   &  0.05  \\
Ir3       &   0.08   &   0.06   &  0.14   &  0.00  \\
Ir4       &   0.29   &   0.00   &  0.15   &  0.18  \\
\end{tabular}
\end{ruledtabular}
\caption{\label{Table_magnetism}
 Relative total energies obtained in GGA (in meV/Fe-atom), spin moments 
 in the muffin-tin
 spheres of the Fe atoms in unsupported (UBC) and supported bichains,
 as well as for Ir surface atoms, and total moments in the
 unit cell (in $\mu_B$) for calculations with 9 and 7 layers of the
 $(5\times1)$ reconstructed Ir(001) substrate. The Ir surface atoms
 are denoted as in Fig.~\ref{fig:sketch}.
}
\end{table}

\section{Magneto-crystalline anisotropy}

\begin{figure}[t]
\begin{center}
\vspace{0.6cm}\includegraphics[scale=0.25]{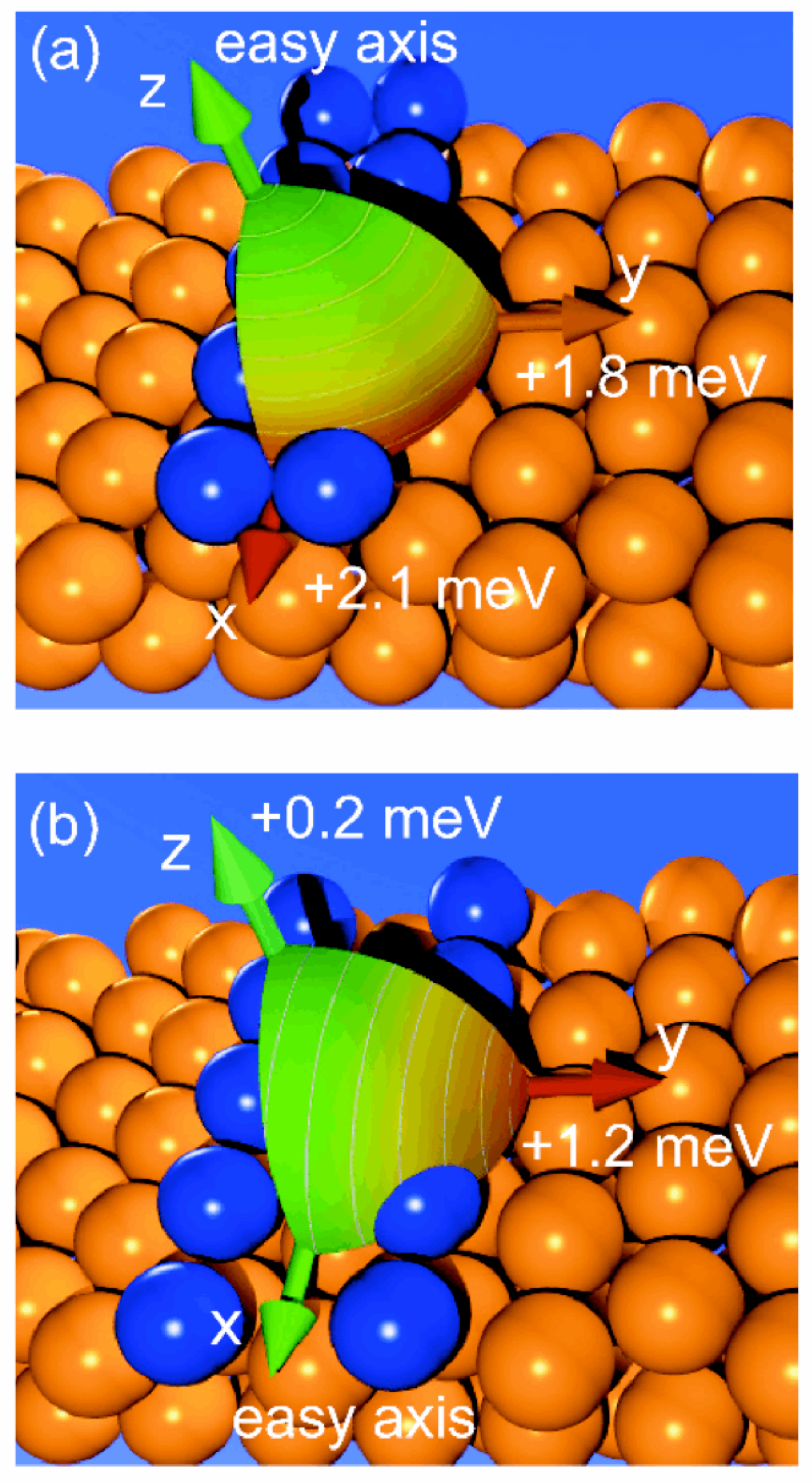}
\end{center}
\caption{\label{fig:MAE} (color online) Sketches of the energy
landscape as a function of magnetization direction for the two
types of chains.
(a) C1-FM configuration with an easy axis
pointing perpendicular to the surface, hard axis along the chain 
(+2.1 meV/Fe atom), and middle axis perpendicular to the
chain and in the surface plane (+1.8 meV/Fe atom). (b) C4-AFM
configuration with an easy axis along the chain axis but only a
small energy difference with respect to the middle axis 
perpendicular to the surface (+0.2 meV/Fe atom), and a hard axis
perpendicular to the chain and in the surface plane (+1.2 meV/Fe atom).
In this plot, we construct the energy landscape based on the lowest
power of the directional cosines of the magnetization with respect to
the crystallographic axes which are allowed by symmetry. The coefficients
are obtained from the values of MAE$_{\Vert}$ and MAE$_{\perp}$.}
\end{figure}

As mentioned in the introduction, the magneto-crystalline anisotropy
energy (MAE) is a key quantity for nanoscale magnets as it determines
the preferred direction of the magnetic moments and is crucial to stabilize
the magnetic order against thermal fluctuations. We have calculated the
MAE for the Fe chains within the GGA employing the force theorem and
found that its value is stable against the chosen ground state upon which
we perform the perturbation. For the C1 structure, we considered the
FM ground state and started from a converged ground state with an
out-of-plane magnetization from which the MAE was obtained by applying
the force theorem for three possible high-symmetry directions:
perpendicular to the surface and the chain axis (along the $z$-axis),
parallel to the axis of the chain ($\Vert$-direction) and in the surface
plane perpendicular to the chain's axis ($\perp$-direction). We define
two principal energies:
MAE$_{\Vert}$ = E$_{\rm tot}$($\Vert$) $-$ E$_{\rm tot}$($z$), and
MAE$_{\perp}$ = E$_{\rm tot}$($\perp$) $-$ E$_{\rm tot}$($z$).
The number of $k$-points used for calculations
constituted 144 in the full 2D-BZ. We carefully tested the MAE values
with respect to the number of $k$-points. Using 144 $k$-points
results in an accuracy of not less than 0.3~meV/Fe. We find that in the
C1-FM configuration the calculated values are 2.1~meV/Fe for
MAE$_{\Vert}$ and 1.8~meV/Fe for MAE$_{\perp}$ which corresponds to a
magnetization along the $z$-axis in the ground state, i.e.~perpendicular
to the surface and chain axis, see Fig.~\ref{fig:MAE}(a).

For the calculations of the MAE in the C4-AFM configuration we used
a slab of 7 layers of the $(5\times1)$ Ir(001) substrate with 8 Fe
chain atoms in the unit cell and 144 $k$-points in the full 2D-BZ
(again the stability of the MAE with respect to the number of
$k$-points was carefully checked). We obtain values of $-$0.2~meV/Fe
for MAE$_{\Vert}$ and 1.2~meV/Fe for MAE$_{\perp}$ corresponding
to a ground state with the magnetic moments along the chain axis,
see Fig.~\ref{fig:MAE}(b). Considering the accuracy of our
calculations we conclude that the easy axis of the magnetization
for the Fe biatomic chains in the C4-AFM configuration
constitutes a plane which cuts through the chain axis and is perpendicular
to the substrate. 

The MAE and its dependence on the configuration can be qualitatively
related to the anisotropy of the orbital moments in the system.\cite{Bruno}
For this purpose for different magnetization directions we compare the
orbital moments inside the atomic spheres of Fe chain atoms, $\mu_L^{\rm Fe}$,
of the Ir atoms in the surface layer, $\mu_L^{\rm Ir}$, and the total orbital
moment per unit cell, $\mu_L^{\rm tot}$, defined as the sum over the moments 
of the two Fe chain atoms and the five Ir surface atoms 
on one side of the slab. For the evaluation of $\mu_L^{\rm tot}$ we do not take
into account the much smaller contributions from Ir atoms deeper inside the slab.
For the antiferromagnetic ground state in the C4 geometry, one should note
that the sign of the orbital moments switches for atoms of antiparallel
magnetization. Therefore, we define $\mu_L^{\rm tot}$ in this case as the
orbital moment summed over atoms in one half of the $(2\times5)$ unit cell
on one side of the slab.

In the C1-FM state the anisotropy of $\mu_L^{\rm tot}$ is in qualitative
agreement with the anisotropy of the total energy: while for an out-of-plane
magnetization $\mu_L^{\rm tot}$ reaches a value of 0.126$\mu_B$, it constitutes
only 0.083$\mu_B$ and 0.035$\mu_B$ for the $\perp$- and $\Vert$-directions of
the magnetization, respectively. Therefore, the easy axis coincides with that
along which the orbital moment is largest.\cite{Bruno} This anisotropy of
$\mu_L^{\rm tot}$ can be explained based on the dependence of the Ir
contributions on the magnetization direction. For an out-of-plane magnetization,
the values of $\mu_L^{\rm Ir}$ are small and the total orbital moment is
dominated by the Fe atoms. In contrast, for a magnetization along the $\perp$- 
and the $\Vert$-direction the Ir orbital moments, $\mu_L^{\rm Ir}$, reach
significant values, however, of opposite sign with respect to $\mu_L^{\rm Fe}$.
For the $\perp$-magnetization direction with $\mu_L^{\rm Fe}=0.077\mu_B$ the
value of the orbital moment of the Ir4 atom (see Fig.~2) is $-$0.013$\mu_B$,
while for the $\Vert$-magnetization it even reaches $-$0.029$\mu_B$ and the
corresponding value of $\mu_L^{\rm Fe}$ is only 0.058$\mu_B$.

In the C4-AFM configuration of the Fe chains the agreement between the anisotropy
of the orbital moment and of the total energy is even better. While for the
$\perp$-magnetization direction the value of $\mu_L^{\rm tot}$ is 0.160$\mu_B$,
it reaches much larger values of 0.239$\mu_B$ and 0.253$\mu_B$ for the $z$- and
$\Vert$-direction, respectively. 
The direction of the smallest total orbital moment, $\mu_L^{\rm tot}$, coincides with
the hard axis and, moreover, a very
small difference between the values of $\mu_L^{\rm tot}$ for the two other directions 
corresponds to a very small energy difference MAE$_{\Vert}$. In this structural
arrangement the anisotropy of the Fe orbital moments, $\mu_L^{\rm Fe}$, dominates the 
anisotropy of $\mu_L^{\rm tot}$. The contribution of the Ir atoms to the total orbital 
moment is nearly independent of the magnetization direction.
For the $z$- and $\Vert$-magnetization directions $\mu_L^{\rm Fe}$ is 0.100$\mu_B$ and 
0.106$\mu_B$, respectively, while it is only $0.060\mu_B$ and thus much smaller for the 
hard $\perp$-magnetization axis.

\section{Simulation of STM experiments}

In the previous sections, we have demonstrated that the magnetic
properties of biatomic Fe chains on the $(5\times1)$ Ir(001)
surface depend crucially on the atomic arrangement of the atoms.
Both easy magnetization direction and magnetic order change upon
displacements of the Fe atoms which changes the interaction
between the two strands of the chain and their hybridization with
the Ir substrate. In order to verify our predictions experimentally,
a technique with a high lateral resolution seems indispensable.
Therefore, we study theoretically the possibility to use
spin-polarized scanning tunneling microscopy (SP-STM) to
resolve the atomic and magnetic structure of these chains.

\begin{figure}
\begin{center}
\includegraphics[scale=0.6]{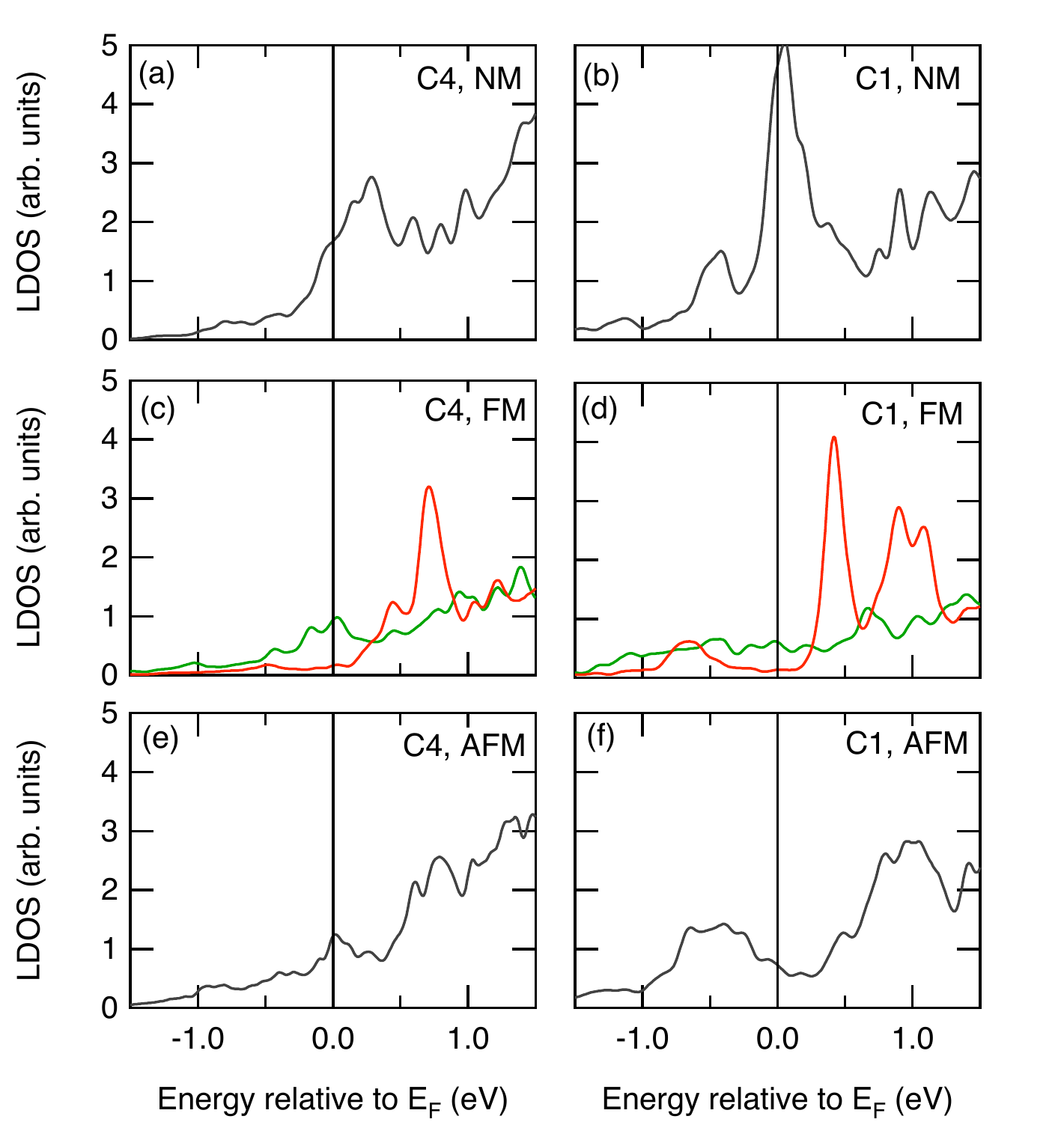}
\end{center}
\caption{\label{fig:VacuumDOS} (color online) 
Local density of states (LDOS) in the vacuum
for the NM, FM and AFM states of Fe biatomic chains on 9 layers of the
$(5\times1)$ reconstructed Ir(001) surface in C4 (left column) and C1 (right
column) configurations. The LDOS is given at a distance of
$\approx$~6.8~{\AA} above the Fe atoms.
 }
\end{figure}

Fig.~\ref{fig:VacuumDOS} displays the calculated local density of
states (LDOS) in the vacuum at a distance of about 6.8~{\AA} from
the Fe chain atoms. Within the Tersoff-Hamann model of STM the
vacuum LDOS is directly comparable with measured $dI/dU$ tunnel
spectra. The comparison of different magnetic states shows distinct
features in the FM solution for both chain structures. Strong peaks
appear in the minority spin channel at +0.7~eV and +0.4~eV in the
C4 and C1 configuration, respectively, see Figs.~\ref{fig:VacuumDOS}(c)
and (d). In the AFM state, we observe a structure of two broad peaks
at $-0.5$~eV and +1.0~eV for the C1 chains, Fig.~\ref{fig:VacuumDOS}(f),
while the C4 chains display a relatively featureless structure with
a small peak at the Fermi energy. Considering the two favorable C1-FM
and C4-AFM states of the chains, it seems that the two types can be
distinguished by the strong peak of the C1-FM configuration.

\begin{figure*}
\begin{center}
\includegraphics[scale=0.25]{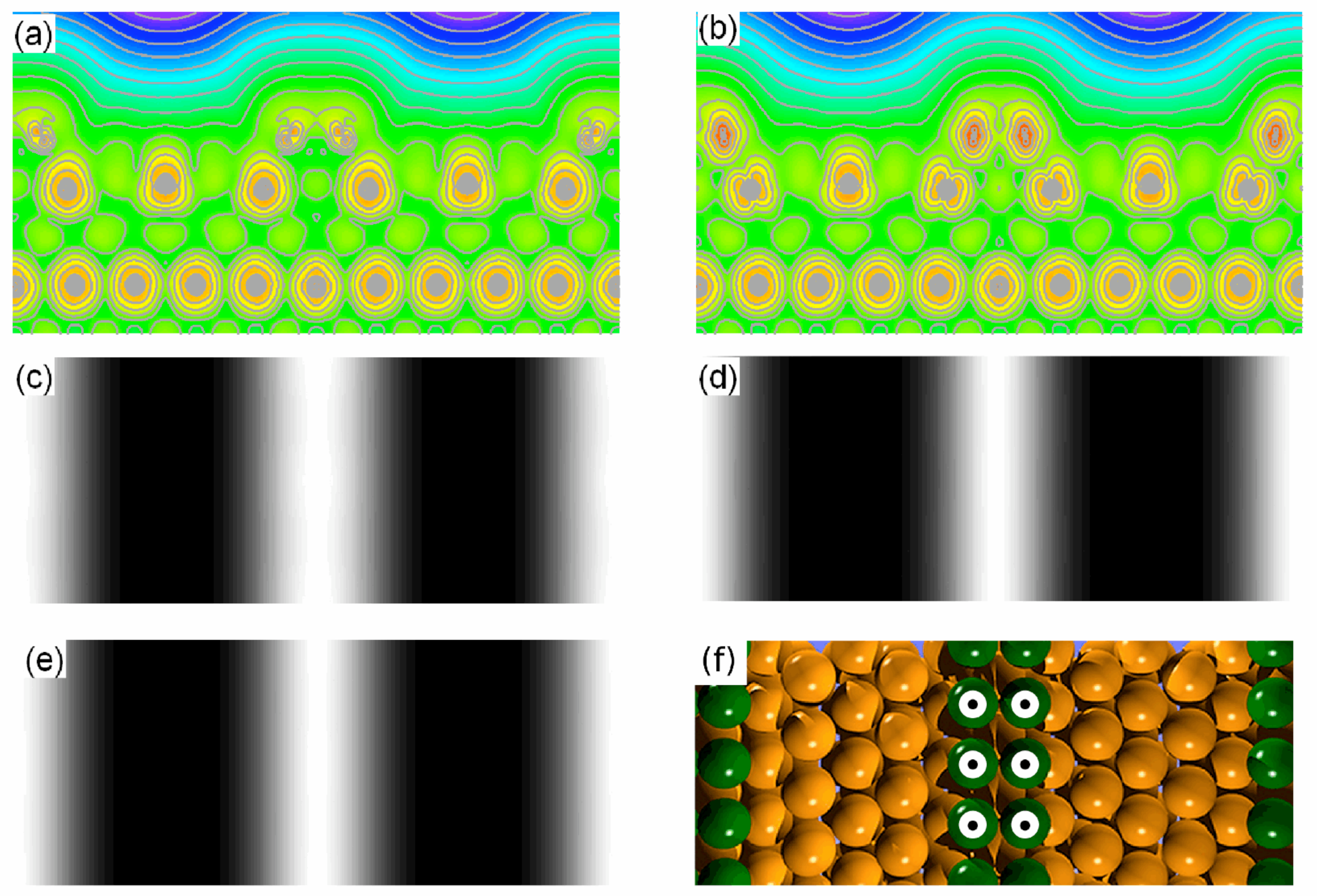}
\end{center}
\caption{\label{fig:STM_C1} (color online) Partial charge density plots for
the FM ground state of the C1 structure (with 7 layers of Ir substrate) 
with an out-of-plane easy
magnetization axis in an energy regime of $(E_F, E_F+0.5)$~eV.
(a) shows a cross-section plot of the majority states
and (b) the minority states up to a distance of 5~{\AA} from the Fe chains
 (c.f.~Fig.~\ref{fig1}). (a) and (b) cut through the middle green Fe atoms in (f). 
STM images at a distance
of 5~{\AA} above the Fe chains are shown in (c) for majority electrons,
(d) for minority electrons, and in (e) for the sum of both contributions. The
part of the surface displayed in the STM images is given in (f).
Note that the charge density plots are very similar for both spin directions, 
with the width of the charge density around the Fe atoms somewhat smaller 
for minority electrons.}
\end{figure*}

\begin{figure*}
\begin{center}
\includegraphics[scale=0.35]{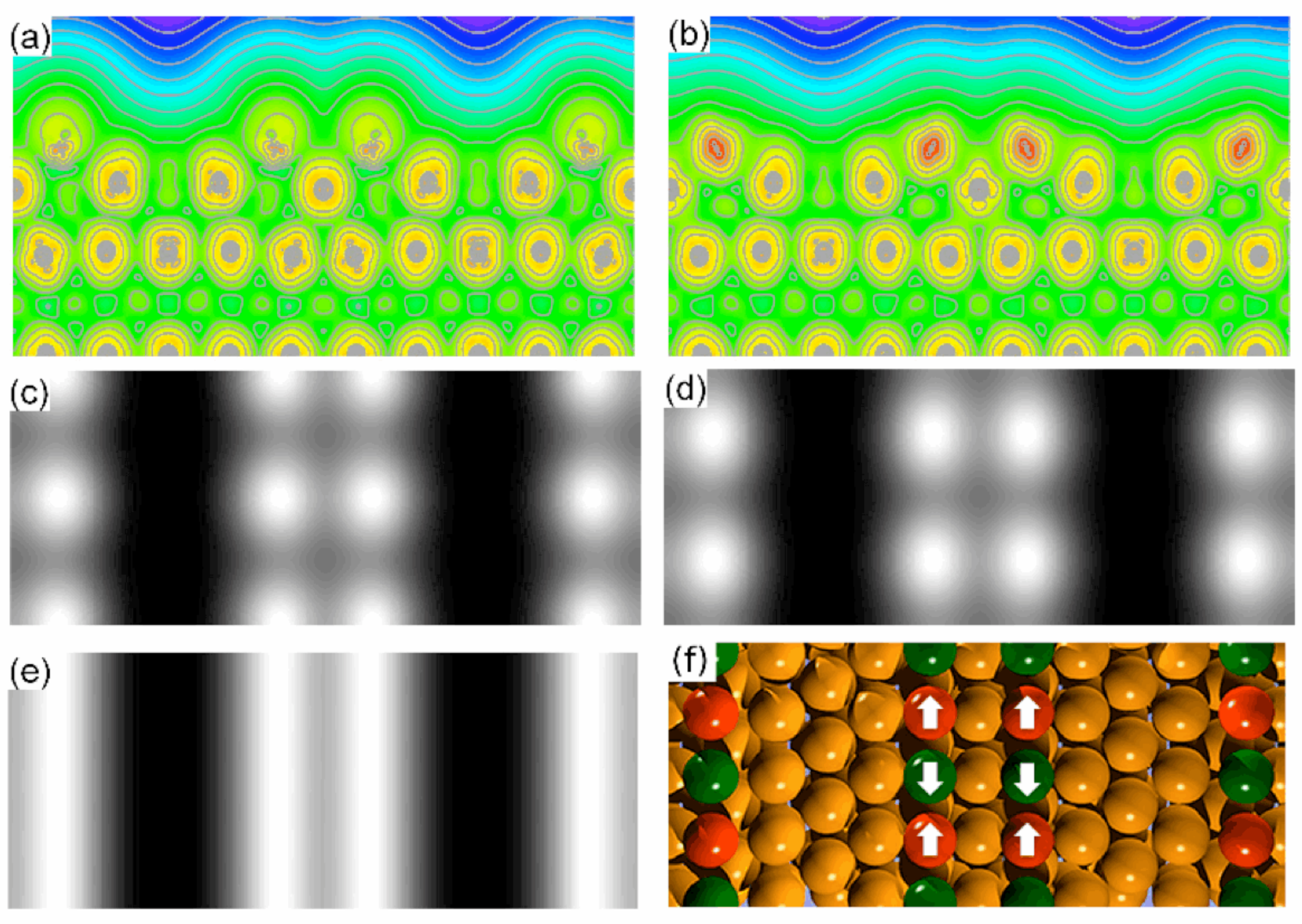}
\end{center}
\caption{\label{fig:STM_C4} (color online) Partial charge density plots
for the AFM ground state of the C4 structure (with 7 layers of Ir substrate)
with an easy magnetization axis along the chains in an energy regime of
$(E_F, E_F+0.3)$~eV. (a) shows a cross-section plot of the majority states and
(b) the minority states up to a distance of 5~{\AA} from the Fe chains
(c.f.~Fig.~\ref{fig1}). (a) and (b) cut through the middle green Fe atoms in (f)
and majority and minority states are defined with respect to these Fe atoms.
STM images at a distance of 5~{\AA} above the Fe chains are shown in (c)
for majority electrons, (d) for minority electrons, and in (e) for the sum
of both contributions. The part of the surface displayed in the STM images is
given in (f) where arrows indicate the direction of the Fe spin moments.
Alternating color of Fe chain atoms and arrows are introduced to emphasize
opposite direction of Fe spin moments. Note, that the spin-averaged local DOS
of the two Fe atoms with opposite spin moments is identical which leads to the 
stripe pattern in (e).}
\end{figure*}

A direct way of verifying the magnetic structure of the Fe chains might
be feasible by imaging them in the topography mode of SP-STM. In addition,
the structural arrangement of the atoms might be detectable. Our
simulations of STM and SP-STM images are displayed in Figs.~\ref{fig:STM_C1}
and \ref{fig:STM_C4} for the FM ground state of the C1 chains and the
AFM ground state of the C4 chains, respectively. We have chosen an
energy window corresponding to the unoccupied states close to the
Fermi energy, but the occupied states lead to very similar results.

The two strands of the biatomic Fe chains in the C1 structure are
only 2.35~{\AA} apart which makes it impossible to resolve them in
cross-sectional scans as seen in Fig.~\ref{fig:STM_C1}(a) and (b)
for both spin channels and also in their summation. The corrugation
amplitudes which we calculate from these plots are 1.9 and 2.5 {\AA}
and the apparent width (full width at half maximum
using the topmost lines of constant charge density) amounts to 7.7
and 6.9 {\AA} for the majority and minority spin channel, respectively.
Interestingly, the chains appear wider if majority states of the chain
are imaged than for the minority states which can be explained based
on the orbital character of the dominating states. From the DOS of
the Fe atoms, Fig.~\ref{fig:MT-DOS}(e), we see that the minority
channel is dominated by $d$-electrons while the majority $d$-band
is far below the Fermi energy and therefore, $s$- and $p$-states
provide a large contribution. The $d$-character of the states for
spin down electrons is clearly visible in the cross-sectional plot,
Fig.~\ref{fig:STM_C1}(b), and leads to a sharper image of the chains.
The more delocalized $s$- and $p$-states dominate the majority spin
channel, Fig.~\ref{fig:STM_C1}(a), and lead to a larger apparent
width and smaller corrugation amplitude of the chains.

In the C4 configuration, the Fe chain atoms are much further apart in
the perpendicular direction, $d=4.23$~{\AA}, which is large enough to
allow the resolution of the two strands as can be seen in our simulations
of STM images, shown in Fig.~\ref{fig:STM_C4}. The corrugation amplitude
of the C4-AFM chains in both spin channels is 1.5 and 2.2 {\AA} in the
spin up and down channel, respectively, the corrugation between the two
Fe chain atoms is 0.15 and 0.3 {\AA}, and the apparent widths are about
10.5 {\AA}, i.e.~significantly wider than the C1 FM chains. With a
spin-polarized STM it should further be possible to resolve the
antiferromagnetic spin alignment along the chains as can be seen from
the STM images for the two separate spin channels as seen from
Figs.~\ref{fig:STM_C4}(c) and (d).  For a spin-polarization of the tip
of $P_t=(n_{\uparrow}-n_{\downarrow})/(n_{\uparrow}+n_{\downarrow})= 0.4$,
where $n_{\uparrow}$ and $n_{\downarrow}$ are the majority and minority spin 
LDOS of the tip at the Fermi energy, we obtain a corrugation amplitude of
$\Delta z=0.15$~{\AA} along the chain,~i.e.~the maximum height change as
the tip scans along the chain.

\section{Conclusions}

We performed extensive first-principles calculations to elucidate the
interplay of structure and magnetism in biatomic Fe chains on the
$(5\times1)$ reconstructed Ir(001) surface. We find a crucial influence
of the hybridization of Fe chains with the Ir substrate on the magnetic
ground state of the wires. Depending on the particular structural
arrangement, the magnetic ground state switches from
along-the-chain ferromagnetic for the C1 configuration with a smaller
($\approx$~2.4~\AA) distance between the two strands, to antiferromagnetic
for the C4 state for which this distance constitutes an almost twice larger
value ($\approx$~4.2~\AA). In the C4 configuration, the two strands of the
chain are nearly decoupled in terms of exchange interaction, while we find
strong ferromagnetic coupling in the C1 configuration. We also find that
the direction of the magnetization in these two configurations is different:
while in C1-FM chains the Fe spin moments point out of plane with
a value of the magnetic anisotropy of $\approx$~2~meV/Fe with respect to
in-plane directions, the magnetization in C4-AFM chains can freely rotate
in the plane of along-the-chain and out-of-plane directions at sufficiently
small temperatures, protected from switching to the in-plane perpendicular 
to the chain direction by a value of $\approx$~1.2~meV/Fe.

The two different magnetic types of chains are very close in
total energy, which facilitates their experimental observation and provides
a considerable challenge for experimentalists to verify their magnetic
ground state.  With our calculations we provide theoretical evidence
for the feasibility to use spin-polarized STM to resolve the atomic
arrangement and magnetic order. Considering the rather small calculated
total energy differences between the different magnetic collinear solutions
of the order of 20$-$30~meV/Fe in both types of chains, we also cannot
exclude the occurrence of noncollinear magnetic states either due to
exchange interactions or due to the Dzyaloshinskii-Moriya interaction
driven by spin-orbit coupling.\cite{Bode2007,Ferriani2008} Such
non-collinear calculations were beyond the scope of the present work
due to the large supercell required for realistic modeling of chains
on the $(5\times1)$ reconstructed Ir(001) surface. However, future
investigations of this system need to address this open question.
Qualitatively and quantitatively different energy landscapes of the
magnetization direction in real space could also result in a different
response of the magnetization with respect to external magnetic field
or temperature, providing an additional channel for tackling the
magnetism in these two types of chains experimentally.

Financial support of the Stifterverband f\"ur die Deutsche Wissenschaft
and the Interdisciplinary Nanoscience Center Hamburg are gratefully
acknowledged. We would like to thank Matthias Menzel, Kirsten von Bergmann,
Andr\'e Kubetzka, Paolo Ferriani, Gustav Bihlmayer, Stefan Bl\"ugel and
Roland Wiesendanger for many fruitful discussions.

\end{document}